\documentstyle[aps,prl,epsf,floats]{revtex}
\def\figuresize{\ifpreprintsty 10cm \else 8cm \fi}

\begin{document}

\ifpreprintsty \else
\twocolumn[\hsize\textwidth\columnwidth\hsize\csname@twocolumnfalse%
\endcsname \fi

\draft
\title{Quantum Coulomb glass -- Hartree-Fock approximation versus
      exact diagonalization}    
\author{Frank Epperlein and Michael Schreiber}
\address{Institut f\"{u}r Physik, Technische Universit\"{a}t,
D-09107 Chemnitz, F.\ R.\ Germany}
\author{Thomas Vojta}
\address{Materials Science Institute, University of Oregon, Eugene, OR 97403, USA}
\date{\today}
\maketitle

\begin{abstract}
We investigate the behavior of disordered interacting
electrons in the insulating regime. Our study is based on
the quantum Coulomb glass model which is obtained from the
classical Coulomb glass by adding hopping matrix elements between
neighboring sites. We use  two different numerical 
methods, viz. a Hartree-Fock approximation and an exact diagonalization
and compare the results for the tunneling density of states and the
localization properties in order to determine the range of validity of the
Hartree-Fock method. We find that the Hartree-Fock method
gives a good approximation for the density of states for all energies 
but represents the localization properties correctly 
close to the Fermi level only. 
Some consequences for the localization of disordered interacting
electrons are discussed.
\end{abstract}

\ifpreprintsty \else
] \fi              


The physics of  disordered interacting electrons has been a subject of great
interest within the last two decades. Most work, both experimental and
theoretical has concentrated on the metallic regime where experiments are
easier to carry out and theoretical studies can be based on established
perturbative methods \cite{reviews}.
In comparison, the insulating regime has seen much less activities. 
Experimentally, this is due to the fact that transport properties vanish
in an insulator at zero temperature.
On the theoretical side, the main reason is that perturbative methods 
cannot be applied since the insulating (classical) limit 
\cite{pollak70,es75} itself is a complicated 
many-particle problem. The
prototype model in the classical insulating regime is the
Coulomb glass
model \cite{cgrev} which describes the electrons as classical point
charges. 

Recently, the attention has focused on the quantum insulating regime closer to the 
metal-insulator transition (MIT) where the description in terms of
classical point charges becomes questionable. These new studies try to
address two main problems:
(i) one wants to gain an understanding of the MIT itself by approaching it 
from the insulating side which is complementary to the usual approach
based on perturbation theory around the metallic system; (ii) one wants to know 
whether the key properties of the insulating limit as, e.g., the Coulomb gap
in the single-particle density of states (DOS) 
\cite{es75}, carry over to the regime close to the MIT where most of the 
experiments on insulators are performed.

The prototype model for the quantum insulating regime is the quantum 
Coulomb glass model \cite{hf,invited,talamantes}.
It is defined on a regular hypercubic lattice with $N=L^d$ ($d$ is the spatial dimensionality) 
sites occupied by $K N$ (spinless) electrons ($0\!<\!K\!<\!1$). To ensure charge neutrality
each lattice site carries a compensating positive charge of  $Ke$.  The Hamiltonian
of the quantum Coulomb glass is 
obtained from the classical Coulomb glass
by adding hopping terms of strength $t$ between nearest neighbor sites.  It reads
\ifpreprintsty
\begin{equation}
H =  -t  \sum_{\langle ij\rangle} (c_i^\dagger c_j + c_j^\dagger c_i) +
       \sum_i (\varphi_i - \mu) n_i + \frac{1}{2}\sum_{i\not=j}(n_i-K)(n_j-K)U_{ij}
\label{eq:Hamiltonian}
\end{equation}
\else
\begin{eqnarray}
H =  &-& t  \sum_{\langle ij\rangle} (c_i^\dagger c_j + c_j^\dagger c_i) \nonumber \\
       &+& \sum_i (\varphi_i - \mu) n_i + \frac{1}{2}\sum_{i\not=j}(n_i-K)(n_j-K)U_{ij}
\label{eq:Hamiltonian}
\end{eqnarray}
\fi
where $c_i^\dagger$ and $c_i$ are the electron creation and annihilation operators
at site $i$, respectively,  and $\langle ij \rangle$ denotes all pairs of nearest 
neighbor sites.
$n_i$ is the occupation number of site $i$ and $\mu$ 
is the chemical potential. The Coulomb interaction $U_{ij} = e^2/r_{ij}$
remains long-ranged since screening breaks down in the insulating phase. 
The random potential values $\varphi_i$ are chosen 
independently from a box distribution of width $2 W_0$ and zero mean.

In a previous paper \cite{hf} we have investigated the quantum Coulomb glass
by means of a Hartree-Fock (HF) approximation. Within this method
the interaction was treated at HF level and the arising self-consistent
disordered single-particle problem was diagonalized numerically. This method
enabled us to study comparatively large systems of up to $10^3$ sites and a
large number of different disorder configurations. We found that the Coulomb gap
persists in the entire insulating phase but becomes narrower when approaching 
the MIT. The depletion of the DOS in turn leads to an enhancement
of localization close to the Fermi level. This enhancement seems to be in 
contradiction to the results of  Talamantes et al. \cite{talamantes} who 
inferred an interaction-induced delocalization from an investigation of the
many-particle spectrum.
Since the HF method is a severe approximation it is not clear whether
the seeming disagreement is an artificial result of the HF approximation
or whether it can be attributed to the use of different criteria \cite{invited}
for  localization in a many-particle system or being in a different part
of parameter space.

In this paper
we therefore investigate the range of validity of the HF approximation
by comparing it to the results of numerically exact diagonalization of small
lattices with up to 16 sites.

We first discuss the single-particle (i.e. tunneling) DOS $g(\varepsilon)$. 
In a many-particle system in general the DOS is defined via the single-particle
Greens function 
\begin{equation}
g(\varepsilon) = - \frac 1 \pi ~ {\rm Im} \, {\rm tr} \, G^R(\varepsilon).
\end{equation}
Within the HF approximation
the single-particle energies are simply given by the eigenvalues of the HF Hamiltonian.
A comparison of the single-particle DOS resulting
from the HF and exact calculations for one particular set of parameters in Fig. 
\ref{fig:doscom01} shows good agreement.
\begin{figure}
\epsfxsize=\figuresize
\centerline{\epsffile{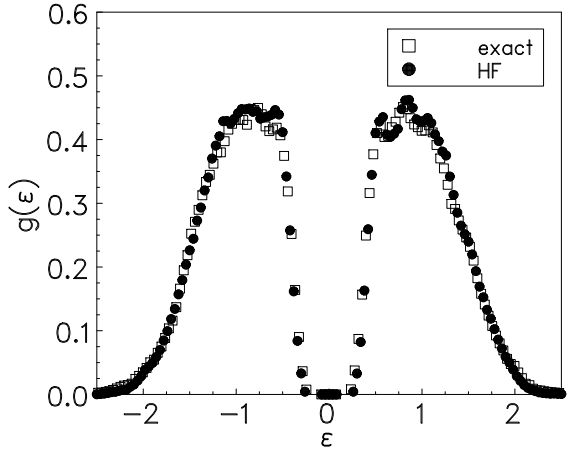}}
\caption{Comparison of the exact and Hartree-Fock results for the
     single-particle DOS of a quantum Coulomb glass with 3x4 sites for
     $W_0\!=\!1$, $K\!=\!0.5$ and $t=0.1$. The data represent averages
     over 1600 different disorder configurations.}
\label{fig:doscom01}
\end{figure}
We have performed analogous calculations for different values of the overlap
($t=0 \ldots 0.8$) and filling factors ($K=0.5$ and 0.25). For all parameter sets
investigated we found that the results of both methods agree within the statistical
errors. Thus we conclude that the single-particle DOS of the quantum Coulomb glass
is well described within the HF approximation. We note, however, that 
we cannot make a statement about the validity of the HF approximation asymptotically
close to the Fermi energy  since the small systems which we can study 
by exact diagonalization always possess a sizeable hard gap which obscures the 
Coulomb gap.

We now turn to the localization properties. The simplest measure of localization
for a single-particle state $|\psi_\nu\rangle$ 
in a non-interacting system is the inverse participation number
\begin{equation} 
P^{-1}_\nu = \sum_j |\langle \psi_\nu | j \rangle |^4
\end{equation}
where the sum runs over all sites $j$.
In practice it is often averaged over all states with a certain energy $\varepsilon$
\begin{equation}
P^{-1}(\varepsilon) = \frac 1 {g(\varepsilon)}\: \frac 1 N \sum_\nu P^{-1}_\nu\: \delta(\varepsilon-\varepsilon_\nu)~.
\label{eq:partnum}
\end{equation}
The generalization of the participation number to many-particle systems is not
straightforward, for a discussion see e.g. Ref.  \cite{invited}. A consistent generalization
of $P^{-1}$ is the probability $R_p$ for a particle to return to its starting site in
infinite time, which can be expressed in terms of single-particle Greens functions
\begin{equation}
R_p(\varepsilon) = \frac 1 N \sum_j \lim_{\delta \to 0} \frac \delta \pi \, G_{jj}(\varepsilon + i \delta)
     \, G_{jj}(\varepsilon - i \delta).
\end{equation}
For non-interacting electrons $P^{-1}(\varepsilon)=R_p(\varepsilon)$. In Fig. \ref{fig:parcom03}
we compare the return probability obtained by Hartree-Fock and exact calculations.
\begin{figure}
\epsfxsize=\figuresize
\centerline{\epsffile{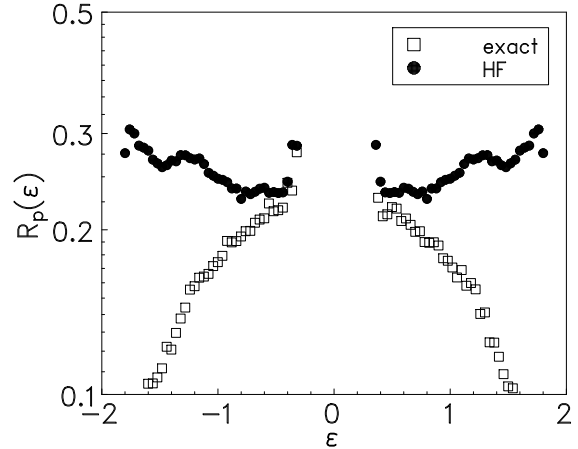}}
\caption{Comparison of the Hartree-Fock and exact results for the return probability 
       $R_p$ for a particle in a quantum Coulomb glass with 3x4 sites for
       $W_0\!=\!1$, $K\!=\!0.5$ and $t=0.3$. The data represent  averages over 1600
 different disorder configurations.}
\label{fig:parcom03}
\end{figure}
Close to the Fermi energy the HF 
and exact results agree reasonably well. Away from the Fermi energy the 
return probability obtained by exact diagonalization decreases drastically which is not 
correctly described
by HF \cite{shklov}. We have carried out analogous calculations for different
values of overlap ($t=0 \ldots 0.8$) and filling factors ($K=0.5$ and 0.25).
For all parameter sets we found the same behavior, viz. the HF method overestimates
the return probability away from the Fermi energy whereas both methods 
agree close to the Fermi level.

To summarize, we have compared the results of a HF approximation
and of an exact diagonalization of the quantum Coulomb glass. We found that 
the results of the two methods for the DOS agree reasonably well for all energies.
The localization properties are correctly described by HF close to the Fermi energy only.
In particular, the interaction-induced enhancement of localization which we found 
\cite{hf} within the HF approximation is also present in the exact calculation.
Consequently, it appears that the seeming contradiction between the HF calculation
\cite{hf} and Ref. \cite{talamantes} is not an artefact of the HF approximation
but either due to the use of {\em different
quantities} to characterize localization in a many-body system
or due to being in {\em different parts} of parameter space. Thus a detailed
investigation of the different localization criteria seems to be  necessary 
\cite{invited}.
The results of the present paper 
not only justify the application of the HF method to analyze
at least some properties of the quantum Coulomb glass 
(taking into account the fact that 
the relevant physics at low temperatures is dominated by excitations close to the
Fermi energy), they also suggest an improved calculational scheme.  It consists
of two steps: (i) solving the model within HF approximation and
(ii) expanding the many-particle states in an energetically cutoff HF basis and
diagonalizing this reduced problem. Such a scheme 
will enable us to treat comparatively
large lattices and still obtain  almost exact results for energies close to the 
Fermi energy. Work along these lines is in progress.

This work was supported in part by the DFG under grant
nos. Schr231/13-1, Vo659/1-1 and SFB393 and by the NSF under grant no. 
DMR-95-10185.


\end{document}